\documentclass[manyauthors]{fundam}

\usepackage{color}
\usepackage{amscd}
\usepackage{amssymb}
\usepackage{times}
\usepackage{graphics}
\usepackage{graphicx}
\usepackage{verbatim}

\usepackage{sim-fi}

\begin{document}
\setcounter{page}{1}
\issue{(submitted)}

\title{Indexing schemes for similarity search:
an illustrated paradigm}

\author{Vladimir Pestov 
\corresponding\\ 
Department of Mathematics and Statistics, \\
University of Ottawa, Ontario, Canada \\
vpest283{@}uottawa.ca \\
http://www.aix1.uottawa.ca/$^\sim$vpest283
\and Aleksandar Stojmirovi\'c \\
Department of Mathematics and Statistics, \\
University of Ottawa, Ontario, Canada\\
astojmir{@}uottawa.ca \\
http://www.aix1.uottawa.ca/$^\sim$astojmir
}
\maketitle
\address{V. Pestov, Department of Mathematics and Statistics, 
University of Ottawa, 
585 King Edward Ave., Ottawa, Ontario, 
Canada K1N 6N5}

\runninghead{V. Pestov, A. Stojmirovi\'c}{Indexing schemes for similarity
search}

\begin{abstract}
We suggest a variation of the Hellerstein---Koutsoupias---Papadimitriou
indexability model for datasets equipped with a similarity measure,
with the aim of 
better understanding the structure of indexing schemes for similarity-based 
search and the geometry of similarity workloads. This in particular provides
a unified approach to a great variety of schemes used to index into metric
spaces and facilitates their transfer to more general similarity measures
such as quasi-metrics. We discuss links between performance of indexing
schemes and high-dimensional geometry.
The concepts and results are illustrated 
on a very large concrete dataset of peptide fragments equipped with a
biologically significant similarity measure.
\end{abstract}


\section{Introduction}

Indexing into very large datasets with the aim of fast
similarity search still remains a challenging and often elusive
problem of data engineering. The main motivation for the present
work comes from sequence-based biology, where high-speed
access methods for biological sequence databases will be 
vital both for developing large-scale datamining
projects \cite{Go} and for testing the nascent
mathematical conceptual models \cite{C-Gr}.

What is needed, is a
fully developed mathematical paradigm of indexability
for similarity search that would incorporate the existing
structures of database theory and possess a predictive power. 
While the fundamental
building blocks - similarity measures, data distributions,
hierarchical tree index structures, and so forth - are in plain view,
the only way they can be assembled together is by examining concrete
datasets of importance and taking one step at a time. 
Theoretical developments and massive amounts
of computational work must proceed in parallel; 
generally, we share the philosophy espoused in
\cite{Papa}.

The master concept was introduced in
the paper \cite{H-K-P} (cf. also \cite{HKMPS02}): a
{\it workload}, $W$, is 
a triple consisting of a search domain $\Omega$, a dataset $X$,
and a set of queries, $\mathcal Q$. An {\it indexing scheme}
according to \cite{H-K-P} is just a collection of blocks covering $X$.
While this concept is fully adequate for many aspects of theory,
we believe that analysis of
indexing schemes for similarity search, with its strong geometric
flavour, requires a more structured approach, and so we put forward a concept
of an indexing scheme as a system of blocks equipped with a tree-like
search structure and decision functions at each step. We also 
suggest the notion 
of a {\it reduction} of one workload to another, allowing one to
create new access methods from the existing ones. 
One example is the new concept of a quasi-metric tree, proposed here.
We discuss how geometry of high dimensions (asymptotic geometric analysis) 
may offer a constructive insight into the nature of the
curse of dimensionality.

Our concepts and results are illustrated throughout on a concrete dataset of
short peptide fragments, containing nearly 24
million data points and equipped with a
biologically significant similarity measure.
In particular, we construct a quasi-metric tree index structure into
our dataset, based on a known idea in molecular biology.
Even if intended as a mere illustration and a
building block for more sophisticated approaches,
this scheme outputs 100 nearest neighbours from the actual dataset
to any one of the $20^{10}$ virtual peptide fragments 
through scanning on average 0.53 $\%$, and at most
3.5 $\%$, of data.

\section{Workloads}
\subsection{Defintion and basic examples} 
A {\it workload} \cite{H-K-P} is a triple
$W=(\Omega,X,{\mathcal Q})$, where $\Omega$ is the {\it domain,} $X$
is a finite subset of the domain ({\it dataset}, or {\it instance}), and 
$\mathcal Q\subseteq 2^{\Omega}$ 
is the set of {\it queries,} that is, some specified subsets of
$\Omega$. {\it Answering a query} $Q\in{\mathcal Q}$ means 
listing all datapoints $x\in X\cap Q$. 

\begin{example}
\label{trivial}
The {\it trivial workload}: $\Omega=X=\{\ast\}$ is a one-element set,
with a sole possible query, $Q=\{\ast\}$. 
\end{example}

\begin{example} Let $X\subseteq\Omega$ be a dataset.
{\it Exact match queries} for $X$
are singletons, that is, sets
$Q=\{\omega\}$, $\omega\in\Omega$.
\end{example}

\begin{example} Let $W_i=(\Omega_i,X_i,{\mathcal Q}_i),i=1,2,\ldots,n$ be 
a finite collection of workloads. Their {\it disjoint sum} is a workload
$W= \sqcup_{i=1}^n W_i$, whose domain is the disjoint union 
$\Omega=\Omega_1\sqcup \Omega_2\sqcup\ldots\sqcup \Omega_n$, the dataset is the disjoint union  
$X=X_1\sqcup X_2\sqcup\ldots\sqcup X_n$, and the queries are of the form
$Q_1\sqcup Q_2\sqcup\ldots\sqcup Q_n$, where $Q_i\in{\mathcal{Q}}_i$,
$i=1,2,\ldots,n$. 
\end{example}
%

\subsection{Similarity queries}
A ({\it dis}){\it similarity measure} on a set $\Omega$ is a
function of two variables $s\colon \Omega\times\Omega\to\R$,
possibly subject to additional properties. A {\it range similarity
query centred at} $x^\ast\in\Omega$ consists of all $x\in\Omega$
determined by the inequality $s(x^\ast,x)<K$ or $>K$, depending on 
the type of similarity measure. 

A {\it similarity workload} is a workload whose queries are
generated by a similarity measure. Different similarity measures,
$S_1$ and $S_2$, on the same domain $\Omega$ can
result in the same set of queries, $\mathcal Q$, in which case we
will call $S_1$ and $S_2$ {\it equivalent}.

Metrics are among the best known similarity measures.
A similarity measure $d(x,y)\geq 0$
is called a {\it quasi-metric} if it
satisfies $d(x,y)=0\Leftrightarrow x=y$ and the triangle inequality,
but is not necessarily symmetric. 
%

\vskip .35cm

\subsection{\label{subsection:illustration}Illustration: short protein fragments} 

 The domain $\Omega$ consists of 
strings of length $m=10$ from the
alphabet $\Sigma$ of 20 standard amino acids:
$\Omega =\Sigma^{10}$. 

The dataset $X$ is formed by all peptide fragments of length 10 
contained in 
the SwissProt database \cite{Bai} of protein sequences of a variety of
biological species (the release 40.30 
of 19-Oct-2002).
The fragments containing parts of low-complexity 
segments masked by the SEG program \cite{Wo}, 
as well as the fragments containing
non-standard letters, were removed. 
The size of the filtered set is $\vert X\vert=23,817,598$
unique fragments (31,380,596 total fragments).

The most commonly used scoring matrix in sequence comparison,
BLOSUM62 \cite{HH}, serves as 
the similarity measure on the alphabet $\Sigma$, and is
extended over the domain $\Sigma^m$ 
via $S(a,b)=\sum_{i=1}^m S(a_i,b_i)$ (the {\it ungapped}
score).

The formula $d(a, b) = s(a,a) - s(a,b)$, $a,b \in \Sigma$,
applied to the similarity measure given by BLOSUM62,
as well as of most other matrices from the BLOSUM
family, is a quasi-metric on $\Sigma$ (Figure \ref{fig:blosum62qd}).
One can now prove that the quasi-metric $\tilde d$ on the domain given by 
$\tilde d(a,b)=\sum_{i=1}^m d(a_i,b_i)$ is equivalent to the
similarity measure $S$.

\begin{figure}
\centering
\scalebox{1}[1]{\tt\tiny
\begin{tabular}{l@{  }c@{}c@{}c@{}c@{}c@{}c@{}c@{}c@{}c@{}c@{}c@{}c@{}c@{}c@{}c@{}c@{}c@{}c@{}c@{}c@{}c@{}}
  & \bf T & \bf S & \bf A & \bf N & \bf I & \bf V & \bf L & \bf M & \bf K & \bf R & \bf D & \bf E & \bf Q & \bf W & \bf F & \bf Y & \bf H & \bf G & \bf P & \bf C \\
\bf T & \gh{0} & \gh{3} & \gh{4} & \gh{6} & 5 & 4 & 5 & 6 & 6 & 6 & 7 & 6 & 6 & 13 &8 & 9 & 10 & 8 & 8 & 10 \\
\bf S & \gh{4} & \gh{0} & \gh{3} & \gh{5} & 6 & 6 & 6 & 6 & 5 & 6 & 6 & 5 & 5 & 14 & 8 & 9 &  9 & 6 & 8 & 10 \\
\bf A & \gh{5} & \gh{3} & \gh{0} & \gh{8} & 5 & 4 & 5 & 6 & 6 & 6 & 8 & 6 & 6 & 14 & 8 & 9 & 10 & 6 & 8 & 9 \\
\bf N & \gh{5} & \gh{3} & \gh{6} & \gh{0} & 7 & 7 & 7 & 7 & 5 & 5 & 5 & 5 & 5 & 15 & 9 & 9 & 7 & 6 & 9 & 12 \\
\bf I & 6 & 6 & 5 & 9 & \gh{0} & \gh{1} & \gh{2} & \gh{4} & 8 & 8 & 9 & 8 & 8 & 14 & 6 & 8 & 11 & 10 & 10 & 10 \\
\bf V & 5 & 6 & 4 & 9 & \gh{1} & \gh{0} & \gh{3} & \gh{4} & 7 & 8 & 9 & 7 & 7 & 14 & 7 & 8 & 11 & 9  & 9  & 10 \\
\bf L & 6 & 6 & 5 & 9 & \gh{2} & \gh{3} & \gh{0} & \gh{3} & 7 & 7 & 10& 8 & 7 & 13 & 6 & 8 &11&10& 10& 10 \\
\bf M & 6 & 5 & 5 & 8 & \gh{3} & \gh{3} & \gh{2} & \gh{0} & 6 & 6 & 9 & 7 & 5 &12 & 6 & 8 &10 & 9 & 9 &10 \\
\bf K & 6 & 4 & 5 & 6 & 7 & 6 & 6 & 6 & \gh{0} & \gh{3} & \gh{7} & \gh{4} & \gh{4} &14&  9 & 9 & 9 & 8 & 8 &12 \\
\bf R & 6 & 5 & 5 & 6 & 7 & 7 & 6 & 6 & \gh{3} & \gh{0} & \gh{8} & \gh{5} & \gh{4} &14&  9 & 9 & 8 & 8 & 9 &12 \\
\bf D & 6 & 4 & 6 & 5 & 7 & 7 & 8 & 8 & \gh{6} & \gh{7} & \gh{0} & \gh{3} & \gh{5} &15&  9 &10 & 9 & 7 & 8 &12 \\
\bf E & 6 & 4 & 5 & 6 & 7 & 6 & 7 & 7 & \gh{4} & \gh{5} & \gh{4} & \gh{0} & \gh{3} &14&  9 & 9 & 8 & 8 & 8 &13 \\
\bf Q & 6 & 4 & 5 & 6 & 7 & 6 & 6 & 5 & \gh{4} & \gh{4} & \gh{6} & \gh{3} & \gh{0} &13&  9 & 8 & 8 & 8 & 8 &12 \\
\bf W & 7 & 7 & 7 &10 & 7 & 7 & 6 & 6 & 8 & 8 &10 & 8 & 7 & \gh{0}&  \gh{5} & \gh{5} &\gh{10} & 8 &11 &11 \\
\bf F & 7 & 6 & 6 & 9 & 4 & 5 & 4 & 5 & 8 & 8 & 9 & 8 & 8 &\gh{10}&  \gh{0} & \gh{4} & \gh{9} & 9 &11 &11 \\
\bf Y & 7 & 6 & 6 & 8 & 5 & 5 & 5 & 6 & 7 & 7 & 9 & 7 & 6 & \gh{9}&  \gh{3} & \gh{0} & \gh{6} & 9 &10 &11 \\
\bf H & 7 & 5 & 6 & 5 & 7 & 7 & 7 & 7 & 6 & 5 & 7 & 5 & 5 &\gh{13}&  \gh{7} & \gh{5} & \gh{0} & 8 & 9 &12 \\
\bf G & 7 & 4 & 4 & 6 & 8 & 7 & 8 & 8 & 7 & 7 & 7 & 7 & 7 &13&  9 &10 &10 & \gh{0} & \gh{9} &\gh{12} \\
\bf P & 6 & 5 & 5 & 8 & 7 & 6 & 7 & 7 & 6 & 7 & 7 & 6 & 6 &15& 10 &10 &10 & \gh{8} & \gh{0} &\gh{12} \\
\bf C & 6 & 5 & 4 & 9 & 5 & 5 & 5 & 6 & 8 & 8 & 9 & 9 & 8 &13&  8 & 9 &11 & \gh{9} &\gh{10} & \gh{0} \\
\end{tabular}
}
\caption{\small BLOSUM62 asymmetric distances. Distances within members
of the alphabet partition used for indexing (cf. Subsect. \ref{illu}
 below) are greyed.}
\label{fig:blosum62qd}
\end{figure}

\vskip .35cm

\subsection{Inner and outer workloads}
We call a workload $W$ {\it inner} if $X=\Omega$, otherwise
$W$ is {\it outer.} Typically, for outer workloads
$\vert X\vert\ll\vert\Omega\vert$. 

\begin{example}
Our illustrative workload is outer, with the ratio
$\vert X\vert/\vert\Omega\vert = 23,817,598/20^{10}\approx
0.0000023$.

Moreover, Fig. \ref{fig:epsnet} shows that an overwhelming number of points
$\omega\in\Omega$ have neighbours $x\in X$ within the 
distance of $\e=25$, which on average indicates high biological relevance. 
For this reason, most of the possible queries
$Q=B_\e(\omega)$ are meaningful, and our illustrative 
workload is indeed outer in a fundamental way.

\begin{figure}
\centering
\scalebox{0.4}[0.4]{\includegraphics{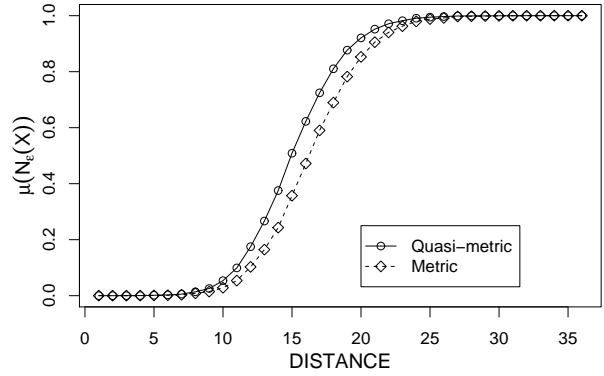}}
\caption{\small Growth with regard to the product measure of 
$\e$-neighbourhoods 
of our illustrative dataset $X$ in $\Omega=\Sigma^{10}$. The $\e$-neighbourhoods
are formed with regard to quasi-metric $d$ (Subsect. 
\ref{subsection:illustration}) and
the smallest metric majorizing $d$ (Ex. \ref{mqm} below).
}
\label{fig:epsnet}
\end{figure}
\end{example}

The difference between inner
and outer searches is particularly significant for similarity searches, and is
often underestimated. 

In theory, every workload $W=(\Omega,X,{\mathcal Q})$ can be replaced
with an inner workload $(X,X,{\mathcal{Q}}\vert_X)$, where 
the new set of queries ${\mathcal{Q}}\vert_X$ consists of sets
$Q\cap X$, $Q\in{\mathcal{Q}}$. However, in practical terms this reduction
often makes little sense because of the prohibitively high
complexity of storing and processing the query sets $Q\cap X$. 

%


\vskip .35cm

\section{Indexing schemes}

\subsection{Basic concepts and examples}
An {\it access method} for a workload $W$ is an algorithm
that on an input $Q\in{\mathcal Q}$ outputs all elements of $Q\cap X$.
Typical access methods come from indexing schemes.

For a rooted finite tree $T$ by $L(T)$ we will denote the set of leaf nodes
and by $I(T)$ the set of inner nodes of $T$. 
The notation $t\in T$ will mean that $t$
is a node of $T$, and $C_t$ will denote the set of all
children of a $t\in I(T)$, while the parent of $t$ will be denoted $p(t)$.

\begin{definition}
Let $W=(\Omega,X,{\mathcal Q})$ be a workload. An {\it indexing scheme}
on $W$ is a triple ${\mathcal I}=(T,{\mathcal B},\F)$,
where 
\begin{itemize}
\item $T$ is a rooted finite tree, with root node $\ast$,
\item ${\mathcal{B}}$
is a collection of subsets $B_t\subseteq\Omega$ ({\it blocks}, or
{\it bins}), where $t\in L(T)$.
\item $\F=\{F_{t}\colon t\in I(T)\}$ 
is a collection of set-valued
{\it decision functions,} $F_t\colon {\mathcal Q} \to 2^{C_t}$,
where each value
$F_{t}(Q)\subseteq C_t$ is a subset of children of the node $t$.
\end{itemize}
\end{definition}

\begin{definition} An indexing scheme ${\mathcal I}=(T,{\mathcal B},\F)$ for a
workload $W=(\Omega,X,{\mathcal Q})$ will be called {\it consistent} if
the following is an access method.
\end{definition}
\begin{algorithm} $\,$
\label{mainal}
\newcommand{\keyw}[1]{{\bf #1}}
\begin{tabbing}
\quad \=\quad \=\quad \=\quad\=\quad\kill
\keyw{on input} $Q$ \keyw{do} \\
\>  set $A_0=\{\ast\}$ \\
\>\keyw{for} each $i=0,1,\ldots$ \keyw{do} \\
\>\>\keyw{if} $A_i\neq\emptyset$ \\
\>\>\keyw{then} for each $t\in A_i$ \keyw{do} \\
\>\>\>\keyw{if} $t$ is not a leaf node \\
\>\> \>\keyw{then} $A_{i+1}\leftarrow A_{i+1}\cup F_{t}(Q)$ \\
\>\>\>\keyw{else} \keyw{for} each $x\in B_t$ \keyw{do}\\
\>\>\>\>\keyw{if} $x\in Q$ \\
\>\>\>\>\keyw{then} $A\leftarrow A\cup\{x\}$\\
\keyw{return} $A$
\end{tabbing}
\end{algorithm}

The following is an obvious and easy to verify sufficient condition for consistency. 

\begin{proposition}\label{consistent} 
An indexing scheme ${\mathcal I}=(T,\mathcal{B},\F)$ for a workload
$W=(\Omega,X,{\mathcal Q})$ is consistent if for every
$Q\in\mathcal{Q}$ and for every $x\in Q\cap X$ there exists
$t\in L(T)$ such that $x\in B_t$ and the path $s_0s_1\ldots s_m$,
where $s_0=\ast$, $s_m=t$ and $s_{i}=p(s_{i+1})$, satisfies
$s_{i+1}\in F_{s_i}(Q)$ for all $i=0,1\ldots m-1$.
%
\end{proposition}

In the future we will be considering consistent indexing schemes only.

\begin{example} A simple linear scan of a dataset $X$ corresponds to
the indexing scheme where $T=\{\ast,\star\}$ has a root and a single
child, $\mathcal B$ consists of a
single block $B_\star=\Omega$, and the decision function $F_\ast$ 
always outputs the same value $\{\star\}$.
\end{example}

\begin{example}
{\it Hashing} can be described in terms of the following indexing
scheme. The tree $T$ has depth one, with its leaves corresponding to bins,
and the decision function $f_{\ast}$ on an input $Q$ outputs the
entire family of bins in which elements of $Q\cap X$ are stored.
\end{example}

\begin{example}
If the domain $\Omega$ is linearly ordered (for instance, assume
$\Omega=\R$) and the set of queries consists
of intervals $[a,b]$, $a,b\in\Omega$, then a well-known and efficient indexing
structure is constructed using 
a binary tree. The nodes $t$ of $T$ can be identified with
elements of $\Omega$ chosen so that the tree is balanced. 
Each decision function $F_t$ on an input $[a,b]$ outputs the set of all
children nodes $s$ of $t$ satisfying 
\[((t-a)(s-a)\geq 0)\wedge ((t-b)(s-b)\geq 0).\]
\label{lodomain}
\end{example}

\begin{remark} The computational complexity of the decision functions
$F_t(Q)$, as well as the amount of `branching' resulting from an application of
Algorithm \ref{mainal}, become major efficiency 
factors in case of similarity-based search, which is why we
feel they should be brought into the picture.
\end{remark}

\subsection{Metric trees}
Let $(\Omega,X,\rho)$ be a similarity workload, where $\rho$ is a metric,
that is, each query $Q=B_\e(\omega)$ is a ball of radius
$\e>0$ around the query centre $\omega\in\Omega$.

A {\it metric tree} is an indexing structure into $(\Omega,X,\rho)$
where the decision functions are of the form
\begin{equation}
F_t(B_\e(\omega))=\{s\in C_t\colon f_s(\omega)\leq\e\}
\label{certif}
\end{equation}
for suitable 1-Lipschitz functions 
$f_s\colon\Omega\to\R$, one for each node $s\in T$. 
(Recall that $f\colon \Omega\to\R$ is 
{\it $1$-Lipschitz} if $\abs{f(x)-f(y)}\leq\rho(x,y)$ for each
$x,y\in \Omega$.) We call those $f_t$ 
{\it certification functions.}
The set $F_t(B_\e(\omega))$ is output by scanning
all children $s$ of $t$ and accepting / rejecting them in accordance with
the above criterion.

\begin{theorem}
\label{mtree}
Let $W=(\Omega,X,\rho)$ be a metric similarity workload. 
Let $T$ be a finite rooted tree,
and let $B_t,t\in T$ be a collection of subsets of $\Omega$ (blocks),
covering $X$ and having the property that 
$X\subseteq\bigcup_{t\in L(T)}B_t\subseteq\Omega$ 
and for every inner node $t$,
$\bigcup_{s\in C_t}(B_s\cap X)\subseteq B_t$.
Let $f_t\colon\Omega\to\R$ be 1-Lipschitz functions with the property
$(\omega\in B_t)\Rightarrow (f_t(\omega)\leq 0)$. Define decision
functions $F_t$ as in Eq. (\ref{certif}).
Then the triple
$(T,\{B_t\}_{t\in L(T)},\{F_t\}_{t\in I(T)})$ is a consistent indexing
scheme for $W$.
\end{theorem}

We omit the proof because a more general result
(Theorem \ref{qmtree}) is proved below.

%

\begin{figure} 
\begin{center}
\scalebox{0.7}{\input{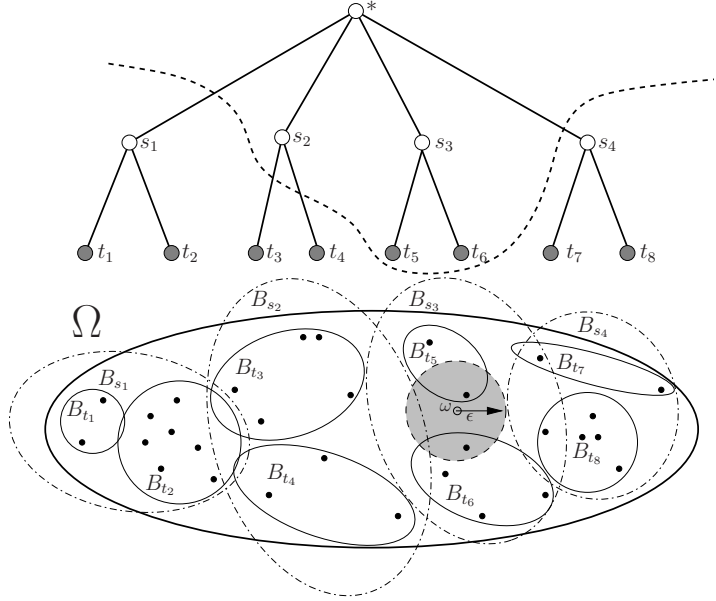}}
\caption[A metric tree indexing scheme.]{A metric tree indexing scheme. 
To retrieve the shaded range query the nodes above the dashed line must be 
scanned; the branches below can be pruned.}
\label{fig:metrictree}
\end{center}
\end{figure}

1-Lipschitz functions $f_t$ with a property required by the assumptions
of Theorem \ref{mtree} always exist.
Once the collection $B_t,t\in T$ of blocks has been chosen, put 
\[f_t(\omega)=\rho(B_t,\omega):=\inf_{x\in B_t}\rho(x,\omega),\]
the distance from a block $B_t$ to an $\omega$. 
However, such distance
functions from sets are typically computationally very expensive. 
The art of constructing a metric tree consists in choosing computationally
inexpensive certification functions that at the same time 
don't result in an excessive amount of branching.

\begin{example} The {\it GNAT} indexing scheme \cite{Brin} uses
certification functions of the form 
\[f_{t_{\pm}}(\omega)= 
\pm\left(\rho(\omega,x_t)-M_t\right),\]
where $x_t$ is a datapoint chosen for the node $t$, $M_t$ is the
median value for the function $\omega\mapsto\rho(\omega,x_t)$, and
$t_{\pm}$ are two children of $t$.
\end{example}

\begin{example} The {\it vp-tree} \cite{Yan}
uses certification functions of the form
\[f_t(\omega)=(1/2)(\rho(x_{t_+},\omega)-\rho(x_{t_-},\omega)),\]
where again $t_{\pm}$ are two children of $t$ and $x_{t_{\pm}}$ are
the {\it vantage points} for the node $t$. 
\end{example}

\begin{example} The {\it M-tree} \cite{CPZ} employs, as certification
functions, those of the form 
\[f_t(\omega)=\rho(x_t,\omega)-\sup_{\tau\in B_t}
\rho(x_t,\tau),\]
where $B_t$ is a block corresponding to the node $t$, $x_t$ is a datapoint
chosen for each node $t$,
and the suprema on the r.h.s. are precomputed and stored.
\label{mtr}
\end{example}

There are many other examples of metric trees, e.g. $k$-d tree, 
{\it gh-}tree, $mvp$-tree, etc. \cite{U,WSB,CNBYM}. 
They all seem to fit into the concept of a general metric
tree equipped with 1-Lipschitz certification functions, 
first formulated in the present exact form in \cite{vp1}.

\begin{example}
\label{ex:hamming}
Suppose $\Omega=X=\{0, 1\}^m$, the set of all binary strings of length $m$. 
The {\it Hamming distance} between two strings $x$ and $y$ is the number of 
terms where $x$ and $y$ differ. A $k$-neighbourhood of any point with respect 
to the Hamming distance can be output by a
combinatorial generation algorithm such as
traversing the binomial tree of order $m$ to depth $k$.
\end{example}

\subsection{Quasi-metric trees}
Quasi-metrics often appear as similarity measures on datasets, and even if
they are being routinely 
replaced with metrics by way of what we call a {\it projective
reduction} of a workload (Ex. \ref{mqm}), 
this may result in a loss of performance (cf. Ex. \ref{qmratio}). 
It is therefore desirable to develop a theory
of indexability for quasi-metric spaces. 

The concept of a 1-Lipschitz function is no longer
adequate.
Indeed, a 1-Lipschitz function $f\colon\Omega\to\R$
remains such with regard to the metric  
$d(x,y)=\max\{\rho(x,y),\rho(y.x)\}$ on $\Omega$, and so using 
1-Lipschitz functions for indexing
in effect amounts to replacing $\rho$ with a coarser metric $d$. 
A subtler concept becomes necessary.
\begin{definition}
Call a function $f$ on a quasi-metric space $(\Omega,\rho)$
{\it left 1-Lipschitz} if for all $x,y\in\Omega$
\[f(x)-f(y)\leq \rho(x,y),\]
and {\it right 1-Lipschitz} if $f(y)-f(x)\leq \rho(x,y)$.
\end{definition}

\begin{example} Let $A$ be a subset of a quasi-metric space $(\Omega,\rho)$.
The distance function from $A$ computed on the left,
$d(x,A)=\inf\{\rho(x,a)\colon a\in A\}$,
is left 1-Lipschitz, while the function
$d(A,x)$ is right 1-Lipschitz.
\end{example}

Now one can establish a quasi-metric (hence more general)
analog of Theorem \ref{mtree}.

\begin{theorem}
\label{qmtree}
Let $W=(\Omega,X,\rho)$ be a quasi-metric similarity workload. 
Let $T$ be a finite rooted tree, and let $B_t,t\in T$ be blocks
covering $X$ in such a way that 
$X\subseteq\bigcup_{t\in L(T)}B_t\subseteq\Omega$ 
and for every inner node $t$,
$\bigcup_{s\in C_t}(B_s\cap X)\subseteq B_t$.
Let $f_t\colon\Omega\to\R$ be left 1-Lipschitz functions such that
$(\omega\in B_t)\Rightarrow (f_t(\omega)\leq 0)$, $t\in I(T)$. Define decision
functions $F_t$ as in Eq. (\ref{certif}).
Then the triple
$(T,\{B_t\}_{t\in L(T)},\{F_t\}_{t\in I(T)})$ is a consistent indexing
scheme for $W$. 
\end{theorem}
\begin{proof}
Let $x\in Q\cap X=B_\e(\omega)\cap X$. By the first covering assumption
above, there exists a leaf node $t$ such that $x\in B_t$. Consider
the path $s_0s_1\ldots s_m$ where $s_0=\ast$, $s_m=t$ and 
$s_{i}=p(s_{i+1})$, from root to $t$. By the second covering assumption
above, for each $i=1,2\ldots m$, we have $(B_t\cap X)\subseteq 
(B_{s_{i}}\cap X)\subseteq  B_{s_{i-1}}$ and hence $x\in B_{s_i}$.
It follows that $f_{s_i}(x)\leq 0$ and, since $f_{s_i}$ is a left 
1-Lipschitz function, we have \[f_{s_i}(\omega)\leq 
f_{s_i}(\omega)-f_{s_i}(x)\leq \rho(\omega,x)\leq\e.\] Therefore,
$s_i\in F_{s_{i-1}}$ and consistency follows by Proposition \ref{consistent}.
\end{proof}

\begin{example} 
\label{qmtr}
Many of the particular types of metric trees generalize
to a quasi-metric setting. For instance, M-tree (Ex. \ref{mtr}) leads to
an indexing scheme into quasi-metric spaces if
the certification functions are chosen as
\[f_t(\omega)=\rho(\omega,x_t)-\sup_{\tau\in B_t}
\rho(\tau,x_t),\]
where $B_t$ and $x_t$ are as in Ex. \ref{mtr}.
\end{example}
%

\subsection{Illustration: a quasi-metric tree for protein fragments
\label{illu}}
Here is a simple but rather efficient implementation of a quasi-metric tree
on our workload of peptide fragments (Subs. 2.3).

Let $\Sigma$, $\Omega=\Sigma^m$, and $d$ be as in Subs. 2.3. 
Let $\gamma$ be a partition of the alphabet $\Sigma$, that is, a finite collection
of disjoint subsets covering $\Sigma$. 
Denote by $T$ the prefix tree of $\gamma^m$, that is, nodes of $T$ are
strings of the form $t=A_1A_2\ldots A_l$, where 
$A_i\in\gamma$, $i=1,2,\ldots,l$,
$l\leq m$, and the children of $t$ are all strings of length $l+1$ having
$t$ as its prefix. To every $t$ as above assign a {\it cylinder subset}
$B_t\subseteq\Omega$, consisting of all strings $\omega\in\Sigma^m$ such
that $\omega_i\in A_i$, $i=1,2,\ldots,l$. 

The certification function $f_t$ for the node $t$ is the
distance from the cylinder $B_t$, computed on the left:
$f_t(\omega):=d(\omega, B_t)$. The value of $f_t$ at any $\omega$
can be computed efficiently using precomputed and stored values of
distances from each $a\in\Sigma$ to every 
$A\in\gamma$. The construction of a quasi-metric tree indexing
into $\Sigma^m$ is accomplished as in Th. \ref{qmtree}.

In our case,
the standard amino acid alphabet is partitioned into five groups 
(Figure \ref{fig:blosum62qd}) based on some 
known classification approaches to
aminoacids from biochemistry. This partition 
induces a partition of $\Omega=\Sigma^{10}$ into $5^{10} = 9,765,625$ 
bins. 

\begin{figure}
\centering
\scalebox{0.4}[0.4]{\includegraphics{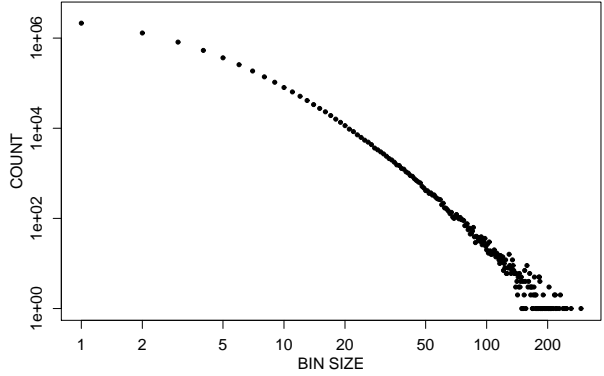}}
\caption{\small Distribution of bin sizes (3,455,126 empty bins
out of 9,765,625 total).}
\label{fig:binsize}
\end{figure}

Since $X$ contains 23,817,598 datapoints, there are on average
2.4 points per bin. 
The actual distribution of bin sizes is strongly 
skewed in favour of small sizes 
(Fig. \ref{fig:binsize}) and appears to follow the DGX distrubition
described in \cite{BFK}. 

\begin{figure}
\centering
\scalebox{0.4}[0.4]{\includegraphics{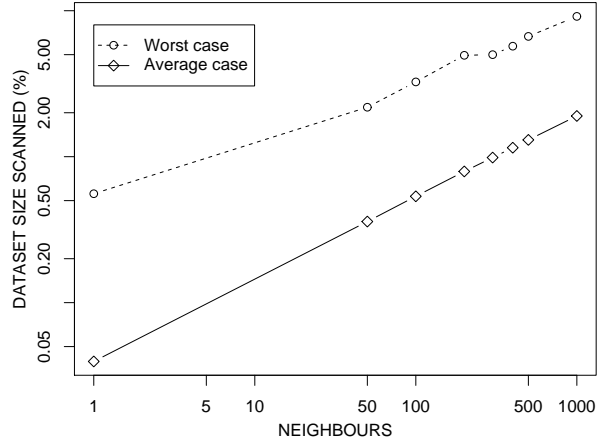}}
\caption{\small Percentage of dataset points scanned to 
obtain $k$ nearest neighbours. Based on 20000 searches for each $k$. 
Query points were sampled with respect to the product measure based 
on amino acid frequencies.}
\label{fig:scanned}
\end{figure}

The performance of our indexing scheme is reflected in Fig.
\ref{fig:scanned}. Recall that an indexing scheme for
similarity search that reduces the fraction of data scanned to below
10 \% is already considered successful. Our figures are many times lower.

\begin{remark}
While other partitions of $\Sigma$ producing different indexing schemes are 
certainly possible, ours can be used for searches based on 
other BLOSUM matrices with little loss of efficiency,
because most amino acid scoring matrices used in practice reflect 
chemical and functional properties of amino acids and hence produce 
very similar collections of queries.
\end{remark}

\section{New indexing schemes from old}
%

\subsection{Disjoint sums}

Any collection of access methods for workloads
$W_1,W_2,\ldots,W_n$ leads to an access method for the disjoint sum 
workload $\sqcup_{i=1}^n W_i$: to answer
a query $Q=\sqcup_{i=1}^n Q_i$, it suffices to answer each query 
$Q_i$, $i=1,2,\ldots,n$, and then merge the outputs.

In particular, if each $W_i$ is equipped with an indexing scheme,
${\mathcal I}_i=(T_i,{\mathcal B}_i,{\mathcal F}_i)$, then
a new indexing scheme for $\sqcup_{i=1}^n W_i$, denoted
${\mathcal I}=\sqcup_{i=1}^n {\mathcal I}_i$, is constructed as follows:
the tree $T$ contains all $T_i$'s as branches beginning at the root node,
while the families of bins and of certification functions for 
$\mathcal I$ are unions of the respective collections
for all ${\mathcal I}_i$, $i=1,2,\ldots,n$.
\vskip .35cm

\subsection{Inductive reduction} 
Let $W_i=(\Omega_i,X_i,{\mathcal Q}_i)$, $i=1,2$ 
be two workloads.
An {\it inductive reduction} of $W_1$ to $W_2$ is a pair of
mappings $i\colon \Omega_2\to \Omega_1$, $i^{\leftarrow}\colon 
{\mathcal Q}_1\to{\mathcal Q}_2$, such that
\begin{itemize}
\item $i(X_2)\supseteq X_1$,
\item for each $Q\in{\mathcal Q}_1$,
$i^{-1}(Q)\subseteq i^{\leftarrow}(Q)$.
\end{itemize}
Notation: $W_2\stackrel{i}{\indred} W_1$.

An access method for $W_2$ leads to an access method for $W_1$,
where a query $Q\in{\mathcal Q}_1$ is answered as follows:

\newcommand{\keyw}[1]{{\bf #1}}
\begin{tabbing}
\quad \=\quad \=\quad \kill
\keyw{on input} $Q$ \keyw{do} \\
\> answer the query $i^{\leftarrow}(Q)$ \\
\>\keyw{for} each $y\in X_2\cap i^{\leftarrow}(Q)$ \keyw{do} \\
\>\>\keyw{if} $i(y)\in Q$ \\
\>\>\keyw{then} add $x=i(y)$ on the list $A$ \\
\keyw{return} $A$
\end{tabbing}

If ${\mathcal I}_2=(T_2,{\mathcal B}_2,{\mathcal F}_2)$ 
is a consistent indexing scheme for $W_2$, then a consistent indexing scheme
${\mathcal I}_1=r_\ast({\mathcal I}_1)$
for $W_1$ is constructed 
by taking $T_1=T_2$, $B^{(1)}_t=
i(B^{(2)}_t)$, and $f^{(1)}_t(Q)=
f^{(2)}_t(i^{\leftarrow}(Q))$ (the upper index $i=1,2$ refers to
the two workloads).

%
%

\begin{example} Let $\Gamma$ be a finite graph of bounded degree,
$k$. Associate to it a
{\it graph workload}, $W_\Gamma$, which is an inner workload with
$X=V_\Gamma$, the set of vertices, and a $k${\it -nearest neighbour} 
query consists in
finding $N$ nearest neighbours of a vertex. 

A {\it linear forest} is a graph that is a disjoint union of paths.
The {\it linear arboricity}, $la(\Gamma)$, 
of a graph $\Gamma$ is the smallest number of 
linear forests whose union is $\Gamma$. 
This number is, in fact, fairly small:
it does not exceed $\lceil 3d/5 \rceil$,
where $d$ is the degree of $\Gamma$ \cite{Alon}. This concept 
leads to an indexing scheme for the graph workload $W_\Gamma$,
as follows. 

Let $F_i$, $i=1,\ldots,la(\Gamma)$ be linear forests. 
Denote $F=\sqcup_{i=1}^{la(\Gamma)}F_i$.
let $\phi\colon F\to \Gamma$ be a surjective
map preserving the adjacency relation.
Every linear forest can be ordered, and indexed into as 
in Ex. \ref{lodomain}. At the next step, 
index into the disjoint sum $F$  as in Subs. 4.1. Finally, index into
$\Gamma$ using the inductive reduction $\phi\colon F\to \Gamma$.
This indexing scheme outputs nearest neighbourhs of any vertex of $\Gamma$
in time $O(d\log n)$, requiring storage space $O(n)$, 
where $n$ is the number of vertices in $\Gamma$.

Of course the similarity workload of the above
type is essentially inner.
\end{example}

\vskip .35cm

\subsection{Projective reduction}
Let $W_i=(\Omega_i,X_i,{\mathcal Q}_i)$, $i=1,2$ be two workloads.
A {\it projective reduction} of $W_1$ to $W_2$ is a pair of mappings
$r\colon \Omega_1\to\Omega_2$,
$r^{\rightarrow}\colon {\mathcal Q}_1\to{\mathcal Q}_2$, such that
\begin{itemize}
\item $r(X_1)\subseteq X_2$,
\item for each $Q\in{\mathcal Q}_1$, $r(Q)\subseteq r^{\rightarrow}(Q)$. 
\end{itemize}

Notation: $W_1 \stackrel{r}{\projred} W_2$.

An access method for $W_2$ leads to an access method for $W_1$,
where a query $Q\in{\mathcal Q}_1$ is answered as follows:

\begin{tabbing}
\quad \=\quad \=\quad \kill
\keyw{on input} $Q$ \keyw{do} \\
\> answer the query $r^{\rightarrow}(Q)$ \\
\>\keyw{for} each $y\in X_2\cap r^{\rightarrow}(Q)$ \keyw{do} \\
\>\>\keyw{for} each $x\in r^{-1}(y)$ \keyw{do} \\
\>\>\keyw{if} $x\in Q$ \\
\>\>\keyw{then} add $x$ on the list $A$ \\
\keyw{return} $A$
\end{tabbing}

Let ${\mathcal I}_2=(T_2,{\mathcal B}_2,{\mathcal F}_2)$ 
be a consistent indexing scheme for $W_2$.
The projective reduction $W_1 \stackrel{r}{\projred} W_2$ canonically
determines an indexing scheme ${\mathcal I}_1=r^\ast({\mathcal I}_2)$
as follows: $T_1=T_2$, $B^{(1)}_t=
r^{-1}(B^{(2)}_t)$, and $f^{(1)}_t(Q)=
f^{(2)}_t(i^{\rightarrow}(Q))$, $i=1,2$.

\begin{example} The linear scan of a dataset 
is a projective reduction to the trivial
workload: $W\projred \{\ast\}$.
\end{example}

If $W=(\Omega,X,{\mathcal Q})$ is a workload and $\Omega^\prime$ is
a domain, then every mapping $r\colon \Omega\to\Omega^\prime$
determines the {\it direct image workload,} 
$r_\ast(W)=(\Omega^\prime,r(X),r({\mathcal Q}))$, where
$r(X)$ is the image of $X$ under $r$ and $r({\mathcal Q})$ is the
family of all queries $r(Q),Q\in{\mathcal Q}$.

\begin{example} \label{ex:blocks} Let $\mathcal B$ be a finite collection 
of {\it blocks} covering
$\Omega$. Define the {\it discrete
workload} $({\mathcal B},{\mathcal B},2^{\mathcal B})$, 
and define the reduction
by mapping each $w\in\Omega$ to the corresponding block and defining
each $\tilde r(Q)$ as the union of all blocks that meet $Q$. 
The corresponding reduction forms a basic building 
block of many indexing schemes.
\end{example}

\begin{example} 
\label{lipschitz}
Let $W_i$, $i=1,2$ be two metric workloads, that is,
their query sets are generated by metrics $d_i$, $i=1,2$. 
In order for a mapping $f\colon \Omega_1\to\Omega_2$ with the property
$f(X_1)\subseteq X_2$ to determine a projective reduction 
$f\colon W_1 \stackrel{r}{\projred} W_2$, it is necessary and sufficient
that $f$ be 1-Lipschitz: indeed, in this case every ball $B_\e^X(x)$
will be mapped inside of the ball $B_\e^Y(f(x))$ in $Y$.
\end{example}

\begin{example} 
Pre-filtering is an often used instance of projective reduction.
In the context of similarity workloads, this normally denotes a procedure
whereby a metric $\rho$ is replaced with a coarser distance $d$ which is
computationally cheaper. This amounts to the 1-Lipschitz map
$(\Omega,X,\rho)\to (\Omega,X,d)$. 
\end{example}

\begin{example}
\label{mqm}
The same applies to quasi-metrics. Moreover,
it is routine to have a quasi-metric, $\rho$, replaced with a metric,
$d$, having the property $\rho(x,y)\leq d(x,y)$, so that one does not miss
any hits. The usual choices are $d(x,y)=\max\{\rho(x,y),\rho(y,x)\}$, or
else $d(x,y)=\rho(x,y)+\rho(y,x)$, followed by a rescaling.
\end{example}

\begin{example} A frequently used tool for
dimensionality reduction of datasets is
the famous Johnson--Lindenstrauss lemma, cf. e.g. \cite{IM} or
Sect. 15.2 in \cite{Mat}.
Let $\Omega=\R^N$ be an Euclidean
space of high dimension, and let $X\subset\R^N$ be a dataset with $n$
points. 
If $\e>0$ and $p$ is a randomly chosen orthogonal projection of $\R^N$ onto a
linear subspace of dimension $k={O(\log n)/\e^2}$, then with overwhelming
probability the mapping $\left(\sqrt{N/k}\right) p$ 
does not distort distances within $X$ by more than the factor of
$1\pm\e$. 

The same is no longer true of 
the entire domain $\Omega=\R^N$, meaning that the technique can be only
applied to indexing for similarity search of the
{\it inner workload} $(X,{\mathcal Q})$, and not the outer workload
$(\Omega,X,{\mathcal Q})$.
\end{example}

\begin{example} A projective reduction of a metric space $\Omega$
to one of a smaller cardinality, $\Omega'$, which in turn is equipped with a
hierarchical tree index structure, is at the core of
a general paradigm of indexing into
metric spaces developed in \cite{CNBYM}.
\end{example}


\subsection{\label{ss:red}Illustration: our indexing scheme}

Our indexing scheme can be also interpreted in terms of 
projective reduction as in example \ref{ex:blocks}.
Denote by $\gamma$ the alphabet consisting of five groups into which
the 20 aminoacids have been partitioned. 
Let $q\colon \Sigma\to \gamma$ be the map assigning to each amino acid the
corresponding group. This map in its turn
determines the map $r=q^m\colon\Omega\to\Omega_\gamma$, where
$\Omega=\Sigma^m$ and $\Omega_\gamma=\gamma^m$. The direct image workload
with domain $\Omega_\gamma$, determined by the map $r$, can be indexed into
using the binomial tree as in example \ref{ex:hamming} 
to generate all bins that can intersect the neighbourhood of the query point.
Denote this indexing scheme by $\mathcal I$. Then the indexing scheme into
$\Omega$, described in Subs. 3.4, is just $r^\ast({\mathcal I})$
as defined in Subs. 4.3.

%
%
%
%

\vskip .35cm

\section{Performance and geometry} 
%


\subsection{Access overhead}
Let $W_i=(\Omega_i,X_i,{\mathcal Q}_i)$, $i=1,2$ be two workloads,
and let $W_1 \stackrel{(r,r^{\rightarrow})}{\projred} W_2$ be a
projective reduction of $W_1$ to $W_2$. The {\it relative access overhead}
of the reduction $r$ is the function $\beta_r\colon {\mathcal{Q}}\to 
[1,+\infty)$, assuming for each query $Q$ the value
$\beta_r(Q):= \abs{r^{-1}\left(r^{\rightarrow}(Q)\right)\cap X}/\abs{Q\cap X}$.

\begin{example} The values for relative access overhead of our indexing
scheme for protein fragments
considered in terms of a projective reduction as in Subs. 4.4 can
be easily obtained from Fig. \ref{fig:scanned}.
\end{example}

\begin{example} 
\label{qmratio}
The access overhead of the projective reduction consisting in 
replacing a quasi-metric
with a metric (Example \ref{mqm}) can be very considerable.
Fig. \ref{fig:qmratio}
shows the overhead in the case of our dataset of fragments, where the
quasi-metric $\rho$ is replaced 
with the metric $d(x,y)=\max\{\rho(x,y),\rho(y,x)\}$.
In our view, this in itself justifies the
development of theory of quasi-metric trees. 
\begin{figure}
\centering
\scalebox{0.4}[0.4]{\includegraphics{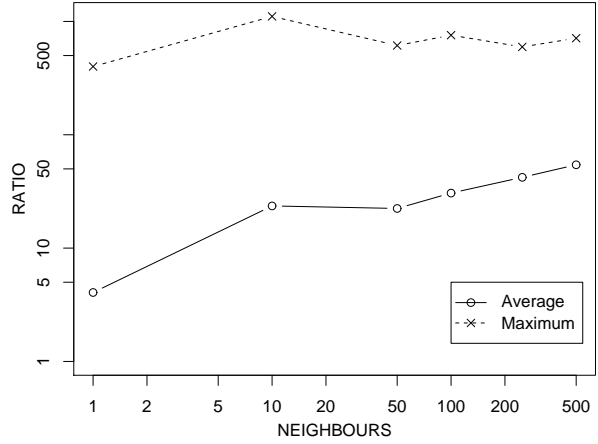}}
\caption{\small Ratio between the sizes of metric and quasi-metric
balls containing $k$ nearest neighbours with respect to quasi-metric.
Each point is based on 5,000 samples.}
\label{fig:qmratio}
\end{figure}
\end{example}

\subsection{Concentration}
Let now $W=(\Omega,X,{\mathcal Q})$ be a similarity workload generated by
a metric, $d$, on the domain. Denote by $\mu$ the normalized 
counting measure supported on the instance
$X$, that is, 
\begin{equation}
\label{mu}
\mu(A)=\abs {A\cap X}/\abs X
\end{equation} 
for an $A\subseteq\Omega$. 
This $\mu$ is a probability measure on $\Omega$. 

The triples of this kind, $(\Omega,\rho,\mu)$, where $\rho$ is a metric
and $d$ is a probability measure on the metric space $(\Omega,\rho)$,
are known as {\it $mm$-spaces,} or {\it probability metric spaces,}
and they form objects of study of {\it geometry of high dimensions} 
({\it asymptotic geometric analysis}),
see \cite{Gr99, Le, Mi} and many references therein.

The central technical concept is that of the
{\it concentration function} $\alpha_\Omega$ of an $mm$-space
$\Omega$: for $\e>0$,
\[\alpha_\Omega(\e)=1-\inf\left\{\mu(A_\e)\colon A\subseteq\Omega,~
\mu(A)\geq\frac 12\right\},\]
and $\alpha_\Omega(0)=\frac 12$. 
If the intrinsic dimension of a triple $(\Omega,\rho,\mu)$ is high, the
concentration function $\alpha_\Omega(\e)$ drops off sharply near zero. 
Typically, the concentration function of a probability metric space
of dimension of order $d$ satisfies the Gaussian estimate
$\alpha_\Omega(\e)\leq C_1\exp(-C_2\e^2d)$, where 
$C_1,C_2$ are suitable constants.
This observation is known as the {\it concentration phenomenon.}

%


The concentration function $\alpha$ is non-increasing, but need not be
strictly monotone. For each $x\geq 0$, denote $\alpha^{\prec}(x)=
\inf\{\e> 0\colon\alpha(\e)\leq x\}$. 
The following result is based on the same ideas as Lemma 4.2 in \cite{vp1}.

\begin{theorem}\label{thm:rngconc}
Let $(\Omega,\rho,\mu)$ be an mm-space, let $\e>0$ and let
$\mathcal{B}$ be a collection 
of subsets $B\subseteq\Omega$ such that $\mu\left( \bigcup\mathcal{B}\right)=1$ 
and for all $B\in\mathcal{B}$, $\mu(B)\leq \xi\leq\frac14$. 
Set $\delta=\alpha^{\prec}(\xi)$. Then, for any $\e>\delta$, 
\begin{enumerate}
\item There exists $\omega\in\Omega$ such that $B_{\e}(\omega)$ meets  
at least \[\min\left\{\left\lceil \frac{1}{2\xi} \right\rceil, 
\left\lceil \frac{1}{\alpha(\e-\delta)}-1 \right\rceil\right\}\]
elements of $\mathcal{B}$.
\item A left ball $B_{\e}(\omega)$ around $\omega\in\Omega$ meets on 
average (in $\omega$) at least 
\[\min\left\{\left\lceil \frac{1}{2\xi}\right\rceil, 
\left\lceil \frac{1}{4\alpha(\e-\delta)} \right\rceil\right\}\] 
elements of $\mathcal{B}$.
\end{enumerate}
\end{theorem}
\begin{proof}
By assumption on each $B\in\mathcal{B}$ and by the choice of $\delta$, 
$\mu(B)\leq \xi\leq\alpha(\delta)$. Decompose $\mathcal{B}$ into a 
collection of pairwise disjoint subfamilies $\mathcal{B}_i$, $i\in I$ 
in a such way that $\alpha(\delta)<\mu(A_i)\leq 2\alpha(\delta)$ for 
each $A_i=\bigcup\mathcal{B}_i$. Clearly, 
\[\frac{1}{2\alpha(\delta)} \leq\frac1{2\xi} \leq\abs{I}<\frac{1}{\alpha(\delta)}.\] 
Let $\delta'=\e-\delta>0$ so that $\left(A_{\delta}\right)_{\delta'}\subseteq A_\e$. 
By Lemma 4.1 of \cite{vp1},  \[\mu\left((A_i)_{\e}\right)\geq 
\mu\left(\left( (A_i)_\delta \right)_{\delta'}  \right)\geq 1-\alpha(\delta'),\] 
and hence the probability that a random left ball of radius $\e$ does not 
intersect $A_i$ is less than \mbox{$\alpha(\e-\delta)$}. For any $J\subseteq I$, 
\[\mu\left( \bigcap_{i\in J}  \left(A_i\right)_\e \right)\geq 1-\abs{J}\alpha(\e-\delta).\] 
The first claim follows by choosing $J$ such that 
$\abs{J}=\min\left\{\abs{I}, 
\left\lceil\frac{1}{\alpha(\e-\delta)}-1\right\rceil \right\}\geq 
\min\left\{\left\lceil\frac{1}{2\xi} \right\rceil, 
\left \lceil\frac{1}{\alpha(\e-\delta)}-1 \right\rceil \right\}$ 
so that $\mu\left( \bigcap_{i\in J} \left(A_i\right)_{\e}\right)>0$. 
To prove the second statement, observe that the probability that a 
random ball of radius $\e$ meets at least 
$\left \lceil \frac{1}{2\alpha\left(\e-\delta\right)} \right\rceil$ elements 
is at least $\frac12$. Hence, the average number of subsets of $\mathcal{B}$ 
intersecting a ball of radius $\e$ is at least 
$\left\lceil \frac{1}{4\alpha\left(\e-\delta\right)} \right\rceil$.
\end{proof} 

This result directly leads to the following corollary stated in terms of 
a range similarity workload (with fixed radius). 

\begin{corollary}\label{cor:rngconc}
Let $\Omega=(W,X,\rho)$ be a metric similarity workload. Suppose the 
dataset $X$ and the query centres are distributed according to the Borel 
probability measure $\mu$ on $\Omega$. Let $\mathcal{B}$ be a finite set 
of blocks such that $\mu(\bigcup\mathcal{B})=1$ and for any $B\in\mathcal{B}$, 
$\mu(B)\leq\xi\leq\frac14$. Then the number of blocks accessed to retrieve 
the query $B_\e(\omega)$, where $\e>\alpha^{\prec}(\xi)$, is on average at 
least $\left\lceil \frac{1}{4\alpha(\e-\alpha^{\prec}(\xi))} \right\rceil$ 
and in the worst case at least 
$\left\lceil \frac{1}{\alpha(\e-\alpha^{\prec}(\xi))}-1 \right\rceil$ or
$\left\lceil \frac{1}{2\xi} \right\rceil$, whichever is smaller. 
\end{corollary}

\begin{example}
In order to apply such estimates to a particular workload, 
one needs to determine its concentration function.
If one equips the dataset of peptide
fragments with a metric as in Ex. \ref{mqm}, then
it is not difficult to derive 
Gaussian upper estimates for the concentration function $\alpha_{W}(\e)$
using standard techniques of asymptotic geometric analysis. 
First, one
estimates the concentration function of $\Omega=\Sigma^{10}$ equipped
with the product measure using the
martingale technique, and then one uses the way $X$ sits inside of
$\Omega$ (the rate of growth of neighbourhoods of the dataset,
cf. Fig. \ref{fig:epsnet}). However, the bounds obtained this way are too
loose and do not lead to meaningful bounds on performance. One needs
to learn how to estimate the concentration function of a workload
more precisely.

Fig. \ref{fig:FSscanned} shows the actual number of bins accessed by
our indexing scheme in order to retrieve $k$ nearest neighbours for 
various $k$. Notice that both the number of bins and the number 
of points of the dataset visited 
(Fig. \ref{fig:scanned}) appear to follow the power law with exponent 
approximately $\frac{1}{2}$ with respect to the number of neighbours retrieved.
\begin{figure}
\centering
\scalebox{0.4}[0.4]{\includegraphics{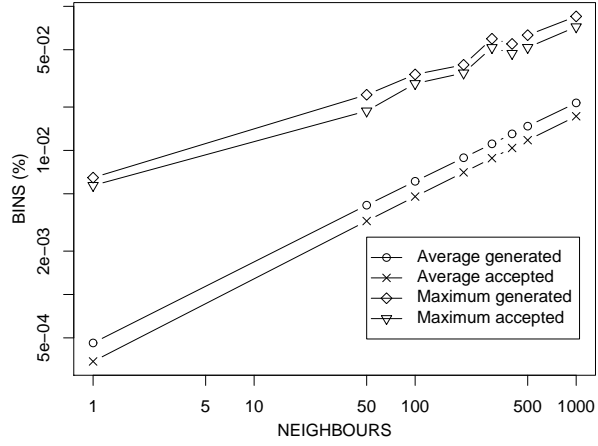}}
\caption{\small Percentage of bins scanned to 
obtain $k$ nearest neighbours. Based on 20000 searches for each $k$. 
The query points were sampled with respect to the product measure based 
on amino acid frequencies.}
\label{fig:FSscanned}
\end{figure}
\end{example}

For a concentual explanation of this phenomenon, consider first the
following example.

\begin{example} 
\label{ex:distexp}
The authors of \cite{ttf} have introduced the
{\it distance exponent} which gives the intrinsic dimension of a metric space
with measure, by assuming that (at least for small $\e$) the size of a ball
$B_\e(x)$ grows proportionally to $\e^N$, where $N$ is the dimension of
the space. (This value is, essentially, an approximation
to the Minkowski dimension of the dataset.) They claimed that performance of
metric trees could be well approximated in terms of the distance exponent.

Fig. \ref{fig:balls} shows (on the log-log scale) 
the rate of growth of measure
of balls $B_\e(\omega)$ in the illustrative dataset of peptide fragments
for the quasi-metric. The rate of growth in the
most meaningful range of $\e$ for similarity search --- and therefore the
distance exponent of our dataset ---
can be estimated as being between 10 and 11. 

\begin{figure}
\centering
\scalebox{0.4}[0.4]{\includegraphics{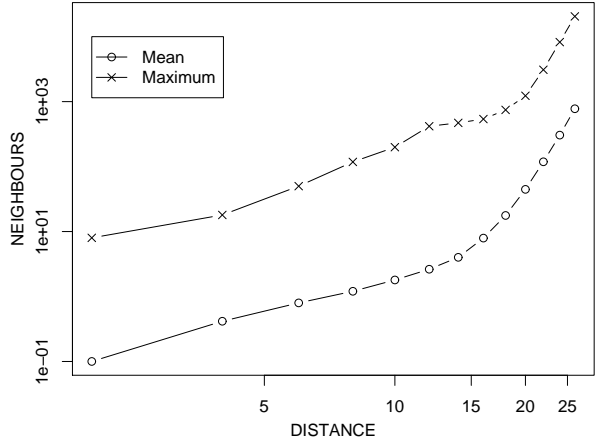}}
\caption{\small Growth of balls in the illustrative dataset.}
\label{fig:balls}
\end{figure}
\end{example}

Returning back to Figure \ref{fig:FSscanned}, clearly the graphs in
question show the average growth of a ball in the projective reduction
$q(\Omega_{\gamma},q(X))$ of our workload (cf. Subs. \ref{ss:red})
against the growth of the ball of the same radius in the
original space $(\Omega,X)$. Denote by $k$ the number of true neighbours
retrieved and by $V(k)$ the corresponding number of fragments scanned.
The power relationship can be written as $V(k) = O(k^{F})$. If we accept
the reasoning behind the distance exponent, that is that $k = O(r^{D})$
where $D$ is the ``dimension'' of the space of protein fragments, it
follows that $V(r) = O(r^{FD})$. Using the same reasoning about the size
of the ball in the reduced workload, we conclude that the ``dimension'' of
it is $FD$, that is, the original dimension $D$ is reduced by the factor
of $F\approx \frac 12$. Assuming that the values of the distance exponent do not depend on
whether a quasi-metric or its associated metric is used and taking the values
of distance exponent estimated in Example \ref{ex:distexp}, the
``dimension'' of the reduced workload 
$(\gamma^m,q(X))$ is somewhere between 5 and 5.5. Thus, our indexing scheme
has reduced the dimension by half.
 
\subsection{Concentration and certification functions}
Let $f\colon\Omega\to\R$ be a 1-Lipschitz function. 
Denote by $M$ the median value of $f$. In asymptotic geometric analysis
it is well known (and easily proved) that
\[\mu\{\omega\colon \abs{f(\omega)-M}>\e\}\leq 2\alpha_\Omega(\e),\]
that is, if $\Omega$ is high-dimensional, the values of $f$ are
concentrated near one value. If one sees such
functions as random variables respecting the distance, the
concentration phenomenon says that on a high-dimensional $\Omega$,
the distribution of $f$ peaks out near one value. 
Using such $f$ as certification functions
in indexing scheme leads to a massive amount of branching and
the dimensionality curse \cite{vp1}.

Yet, there are reasons to believe that the main reason for the
curse of dimensionality is not the inherent high-dimensinality of
datasets, but a poor choice of certification
functions.
Efficient indexing schemes require usage of 
{\it dissipating functions,} that is,
1-Lipschitz functions whose spread of values is more broad, and
which are still computationally cheap.
This interplay between complexity and dissipation is, we believe, at the
very heart of the nature of dimensionality curse.

\begin{example}
One possible reason for a relative efficiency of our quasi-metric tree
may be a good choice of certification functions,
which are somewhat less concentrated than distances from points (Fig. \ref{fig:ddist}).

\begin{figure}
\centering
\scalebox{0.4}[0.4]{\includegraphics{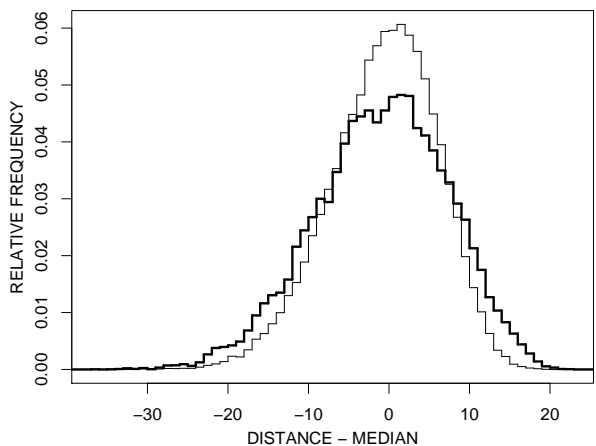}}
\caption{\small Distributions of distances from 40,000 random points to 
a typical point (SEDRELLTEQ) in $\Omega$ and of distances to
a bin (the one containing the above fragment).}
\label{fig:ddist}
\end{figure}
\end{example}

\vskip .35cm

\section{Conclusions}

Our proposed approach to indexing schemes used in similarity search allows
for a unifying look at them and facilitates the task
of transferring the existing expertise to more general similarity measures
than metrics. In particular, we propose the concept of a quasi-metric tree
based on a new notion of left 1-Lipschitz functions, and implement it on
a very large dataset of peptide fragments to obtain a simple yet 
efficient indexing scheme. 

We hope that our concepts and constructions will meld with methods of
geometry of high dimensions and lead to analysis of performance of
indexing schemes for similarity search. 
While we have not yet reached the stage where asymptotic geometric analysis
can give accurate predictions of performance, at least it 
leads to some conceptual understanding of their behaviour.  

We suggest using our dataset of protein fragments 
as a simple benchmark for
testing indexing schemes for similarity search. 

\vskip .35cm

\section{Acknowledgements}
The authors are grateful to Bill Jordan for his
gentle guidance in the area of proteomics and
stimulating discussions. The investigation was
supported by the Marsden Fund of the Royal Society of New Zealand,
by an University of Ottawa start-up grant, and by an NSERC operating grant. 
The second named author
(A.S.) was also supported by a Bright Future PhD scholarship awarded by
the NZ Foundation for Research, Science
and Technology jointly with the Fonterra Research Centre.
\nocite{*}
\bibliographystyle{fundam}
\bibliography{figuide}

\end{document}